\documentstyle[twoside,fleqn,espcrc2]{article}

\newcommand{\AmS}{{\protect\the\textfont2
  A\kern-.1667em\lower.5ex\hbox{M}\kern-.125emS}}
\newcommand{\vierbein}
  {{\footnotesize \left[\!\!\begin{array}{cc}\phi_+&\!\!-\phi_-\\ 
                                        \bar{\phi}_-&\!\!\bar{\phi}_+ 
   \end{array} \!\!\right]}
  }
\newcommand{\vierbeinbar}
  {{\footnotesize \left[\!\!\begin{array}{cc}\bar{\phi}_+&\!\!-\phi_-\\ 
                                        -\bar{\phi}_-&\!\! \phi_+ 
   \end{array} \!\!\right]}
  }
\newcommand{\twoeps}
  { {\footnotesize
    \left[\!\!\begin{array}{c}\epsilon^1 \\ \epsilon^2 \end{array} \!\!\right]}
  }
\newcommand{\twoepstransp}
  { {\footnotesize
   \left[\!\!\begin{array}{cc}-\epsilon^2&\!\!\epsilon^1\end{array} 
   \!\!\right]}
  }
\newcommand{\twoepsconj}
  { {\footnotesize \left[\!\!\begin{array}{c}-\bar{\epsilon}^2 \\ 
    \bar{\epsilon}^1 \end{array}
    \!\!\right] }
  }
\newcommand{\twoepsdag}
  { {\footnotesize \left[\!\!\begin{array}{cc} \bar{\epsilon}^1&\!
          \bar{\epsilon}^2 \end{array}\!\!\right]}
  }
\newcommand{\lambdachi}
  { {\footnotesize
   \left[\!\!\begin{array}{c}\lambda \\ \chi \end{array}\!\!\right]}
  }
\newcommand{\lambdachitransp}
  { {\footnotesize \left[\!\!\begin{array}{cc}-\chi&\!\!\lambda 
    \end{array}\!\!\right]}
  }
\newcommand{\lambdachiconj}
  { {\footnotesize \left[\!\!\begin{array}{c}-\bar{\chi} \\ 
     \bar{\lambda} \end{array}\!\!\right]}
  }
\newcommand{\lambdachidag}
  { {\footnotesize \left[\!\!\begin{array}{cc}\bar{\lambda}&\!\bar{\chi} 
    \end{array}\!\!\right]}
  }
\newcommand{\twopsi}
  { {\footnotesize \left[\!\!\begin{array}{c}\psi_- \\ \psi_+ 
    \end{array}\!\!\right]}
  }
\newcommand{\twopsitransp}
  { {\footnotesize \left[\!\!\begin{array}{cc}-\psi_+&\!\!\psi_- 
    \end{array}\!\!\right]}
  }

\newcommand{\twopsidag}
  { {\footnotesize \left[\!\!\begin{array}{cc}\bar{\psi}_-&\!\bar{\psi}_+ 
    \end{array}\!\!\right]}
  }

\newcommand{\sigmathree}
  { {\footnotesize \left[\!\!\begin{array}{cc}1&\!\\&\!-1 
    \end{array}\!\!\right]}
  }

\title{{\small CPTH-PC496.0297} \\ Partial Spontaneous Breaking of
  Global Supersymmetry\thanks{Research supported in part by the contract
    CHRX-CT93-0340.}\thanks{Contribution to the proceedings of the
  {\em International 
    Symposium Ahrenshoop on the Theory of Elementary Particles},
    Buckow, Germany, August 27-31, 1996.}}

\author{H. Partouche\address{Centre de Physique Th{\'e}orique, 
        Ecole Polytechnique,
        F--91128 Palaiseau c{\'e}dex, France} and B. Pioline$^{a}$}
       
\begin{document}

\begin{abstract}
We review in detail the recently discovered phenomenon of partial 
spontaneous breaking of 
supersymmetry in the case of a $N=2$ pure gauge $U(1)$ theory, and 
recall how the standard lore no-go theorem is evaded. We discuss the extension
of this mechanism to theories with charged matter, and surprisingly find
that the gauging forbids the existence of a magnetic Fayet--Iliopoulos 
term.

\end{abstract}

\maketitle

\section{Introduction}

During the last two years, vast progress has been made in our
understanding of the non-perturbative dynamics of supersymmetric
quantum field theory and string theory. 
$N=2$ supersymmetric theories
present a manifold of vacua (so called moduli space) corresponding to the 
arbitrary values of scalar fields in the flat directions of the potential;
while the physics is determined at a given point on that manifold, 
it admits different (dual) descriptions in each overlapping chart, and
the consistency of the chart changing operations, together with
holomorphy properties often allows for the exact determination of the physics
\cite{SW}, at least at low energies.
On the other hand, flat directions can be lifted in various ways
(gauging of isometries of the moduli space, generation of Fayet--Iliopoulos
terms), and $N=2$ supersymmetry can be spontaneously broken to $N=1$ or $N=0$,
in principle still allowing control of the dynamics of the latter.
Even after partial spontaneous breaking of local extended
supersymmetry had been demonstrated \cite{FGP}, it was so far unanimously believed not to occur
in globally supersymmetric, based on common sense no-go theorems
(recalled in section 2). 
Common sense nevertheless had to retract when an explicit example of
such a phenomenon was found \cite{APT}, namely in the simplest case of
a globally $N=2$ SUSY $U(1)$ pure gauge theory. This model is reminiscent of
the celebrated one considered by Seiberg and Witten
\cite{SW} as an effective theory of a pure $N=2$ $SU(2)$
super-Yang-Mills multiplet Higgsed to $U(1)$, but it uses extra deformations 
compatible with $N=2$ SUSY, but not dynamically generated in the case
of Seiberg--Witten. These Fayet--Iliopoulos terms, introduced in section 3, 
lift the flat directions and leave a $N=1$ supersymmetric vacuum, as
reviewed in section 4. It is then legitimate to ask how general is
this phenomenon. Another surprise comes in section 5, where we try to
extend it to the case of a $U(1)$ theory with charged matter: there appears to
be no such deformation compatible with $N=2$ SUSY anymore.

\section{Partial Breaking no-go Theorem}

The following arguments are general but we use here four dimensional
notations. $N$-extended supersymmetry reads, for 
all $A$ and $B\in \{1,...,N\}$
\begin{equation}
  \left\{Q^A_\alpha , \bar{Q}^B_{\dot{\alpha}} \right\}=H \delta ^{AB} \delta_{\alpha
  \dot{\alpha}} \, ,
\label{QQ}
\end{equation}
and therefore, for {\em any fixed} $A\in \{1,...,N\}$,
\begin{equation}
2H=||\, Q_1^A |0 \rangle \, ||^2+|| \,  Q_2^A |0 \rangle \, ||^2.
\label{H}
\end{equation} 
Now, if it exists $A_0$ such that the vacuum is not annihilated by
the corresponding supersymmetry generator, then the r.h.s. of 
(\ref{H}) 
is strictly positive and so is the Hamiltonian $H$. Then, for {\em any}
other supersymmetry generator $Q^A$, the r.h.s. of (\ref{H}) is also
strictly positive and $Q^A$ is spontaneously broken as
well. So, the no-go theorem states that in a $N$-extended global
supersymmetric theory, either all or no supersymmetry is 
spontaneously
broken. 

This however assumes that the  Hilbert space has a definite positive
metric, which fails in the case of local supersymmetry, and also
that the SUSY charges $Q_\alpha^A$ can be defined at all. It is
a weaker assumption that a SUSY current algebra exists. As was shown
in  \cite{HLP}, the familiar form of the current algebra can be
extended with a $N\times N$ constant matrix $C^{AB}$~:
\begin{eqnarray}
\int d^3 \vec{y} \left\{\bar{J}^B_{0 \dot{\alpha}}(y),J^A_{\mu
  \alpha}(x)  \right\}
  &\!\!\!\! = &\!\!\!\! 2 \sigma ^{\nu}_{\alpha \dot{\alpha}} T_{\nu \mu}(x)
  \delta^{AB} \nonumber \\ \!\!\!\!& &\!\!\!\! +  \sigma_{\mu \alpha \dot{\alpha}} C^{AB} \, .
\label{al}
\end{eqnarray}
The addition of this term is valid since it is only a central
extension, and thus the Jacobi identity is still verified. For the same reason,
the presence of such a  constant term cannot be seen by studying a SUSY
representation in terms of fields.

Now, the point is that it is valid to integrate the l.h.s. of eq. (\ref{al})
over the 3-space variables $\vec{x}$, only when $C^{AB}=0$, since a
constant term would give an infinite
contribution on the other side. Therefore, whenever $C^{AB}\neq 0$,
one cannot write eq. (\ref{QQ}) and the no-go 
theorem breaks down.

\section{N=2, U(1) gauge theory in D=4}

Here we describe the most general N=2 SUSY action for a 
$U(1)$ vector multiplet, using first the formalism of $N=2$ superspace.
We start by considering a  $N=2$ chiral
superfield ${\cal A}$ \cite{GSW,dRcHdWvP}. This multiplet is a
reducible representation of $N=2$, and one can reduce it to obtain a
vector multiplet by imposing the constraint~:
\begin{equation}
(\epsilon_{ij} D^i \sigma_{\mu \nu} D^j)^2{\cal A}=-96\Box \bar{{\cal A}},
\end{equation} 
where the $D^i$ are the $N=2$ supercovariant derivatives. The component fields are then a complex scalar $a$, a $SU(2)_R$ doublet of
chiral fermions $\lambda_i ^\alpha$, a triplet of auxiliary fields
which verifies $\Box \vec{Y}=\Box \vec{Y}^*$, and a self-dual field strength 
now verifying the Bianchi identity, so that it can be traded for
a gauge boson $A_\mu$. 

Now, one may consider a general prepotential $F$ which is a
holomorphic function, and write a $N=2$ Lagrangian
\begin{equation}
{\cal L} =-{\mbox{Im}\over 8\pi}\left[ \int d^2\theta_1 d^2\theta_2 F{\cal
  (A)}-i\vec{e}\cdot \vec{Y} \right] \, ,
\end{equation}
where $\vec{e}$ is a real constant vector of Fayet-Iliopoulos terms
\cite{F},
$\vec{Y}$ being real. 
Instead, for reasons which will become clear in a moment, we 
prefer to start with a {\em non}-reduced $N=2$ chiral superfield ${\cal A}$ 
and the following Lagrangian~:
\begin{equation}
{\cal L}_0 =-{\mbox{Im}\over 8\pi}\int d^2\theta_1 d^2\theta_2 [F({\cal A})-{\cal A}_D {\cal
  A}]
\end{equation}
where ${\cal A}_D$ is a {\it reduced} superfield, i.e. a vector
multiplet. The elimination of this superfield by use of the equations of
motion of its component fields imposes all the reduction relations on
${\cal A}$ which then becomes a vector multiplet as well. Also, the
variation of the superfield ${\cal A}$ gives
\begin{equation}
\left[ F'({\cal A}) - {\cal A}_D \right] \delta {\cal A} = 0 \, ,
\end{equation}
which means that ${\cal A}_D$ is the magnetic dual of the original
electric vector multiplet ${\cal A}$ \cite{SW}.   

Now, the supersymmetric variation of the auxiliary fields  
$\vec{Y}_D$ of ${\cal A}_D$ are total
derivatives. In addition, the variations of $\vec{Y}$ become total derivatives
when the reduction conditions are imposed, which is the case when 
${\cal A}_D$ is eliminated. Therefore, we can consider the previous
Lagrangian to which we add linear terms in $\vec{Y}$ and $\vec{Y}_D$~: 
\begin{equation}
{\cal L}_{\mbox{\em tot}} = {\cal L}_0 + {\cal L}_{F.I.}\,
\mbox{, where}
\end{equation}
\begin{displaymath}
{\cal L}_{F.I.} = {\mbox{Im}\over 8\pi}
\left[ i(\vec{e}+i\vec{p})\cdot \vec{Y}+i\vec{m}\cdot \vec{Y}_D \right] 
\end{displaymath}
and $\vec{e}$, $\vec{p}$, $\vec{m}$ are real constants. Rewriting ${\cal
  L}_{\mbox{\em tot}}$ in component fields and eliminating the dual ones, one can
easily see that
\begin{eqnarray}
{\cal L}_{\mbox{\em tot}}\!\!\!\! &=& \!\!\!\!-{\mbox{Im}\over 8\pi}\left[ \int
  d^2\theta_1 d^2 \theta_2 F({\cal A}_{\vec{m}}) -i\vec{e}\cdot
  \vec{Y}_{\vec{m}}\right] \nonumber \\
\!\!\!\!& &\!\!\!\!-{\vec{p}\cdot\vec{m}\over 8\pi} \, ,\label{ltot} 
\end{eqnarray}
where ${\cal A}_{\vec{m}}$ is the original field ${\cal A}$
where the reducing condition is imposed and in addition $\vec{Y}$ is
replaced by $\vec{Y}+i\vec{m}$. The point is that the addition of such
imaginary constants to the real auxiliary fields $\vec{Y}$ is still
compatible with the reduction condition $\Box \vec{Y}= \Box
\vec{Y}^{*}$. Since the addition of a real constant to $\vec{Y}$ is
the result of the well known
Fayet-Iliopoulos terms $\vec{e}$ for the electric $U(1)$ vector multiplet ${\cal A}$, $\vec{m}$ well deserves the name of
{\em magnetic} Fayet-Iliopoulos term. Notice that the
parameters $\vec{p}$ only appear as a constant term in
eq. (\ref{ltot}). As a result, we can set it to zero 
without loss of generality.

In order to make contact with the more familiar $N=1$ superspace
formalism, we have to choose a particular direction in the $SU(2)_R$
automorphism group of $N=2$. A suitable choice puts $\vec{e}$ and
$\vec{m}$ in the form
\begin{equation}
\vec{e}=(0,e,\xi) \; \; , \; \; \vec{m}=(0,m,0) \, .
\end{equation}
Then, we can rewrite
\begin{equation}
{\cal L}_{\mbox{\em tot}} = {\mbox{Im}\over4\pi}  \left[{\cal L}_0^{1} +  {\cal
    L}_{F.I.}^{1}\right]\, , \, \mbox{where} 
\label{Ltot1}
\end{equation}
\begin{displaymath}
{\cal L}_0^{1} = \int d^2\theta d^2
  \bar{\theta} F'(A)\bar{A} + \int d^2 \theta \frac{1}{2} \tau(A)
  W^\alpha W_\alpha  \, ,
\end{displaymath}
\begin{displaymath}
{\cal L}_{F.I.}^{1} = i \int d^2 \theta \left[ eA+mF'(A)\right] -i \sqrt2\int d^2\theta
  d^2\bar{\theta} \xi V 
\end{displaymath}
and $V$ is a $N=1$ vector
superfield whereas $A$ is a $N=1$ chiral superfield those degrees of
freedom are $A_\mu , \, \lambda_1\equiv \lambda$ and $a, \,
\lambda_2\equiv \chi$, respectively.
Here, we use the notations and conventions of \cite{WB} and \cite{SW}
where $\tau = F''$. In that case, the fermionic mass matrix and scalar
potential take the form~: 
\begin{displaymath}
{\cal M}={\tau'\over 32\pi\tau_2} 
             \lambdachitransp  
{{\footnotesize \left[\!\!\begin{array}{cc}-\xi&\!\!i(e+m\bar{\tau})\\ 
                                        -i(e+m\bar{\tau})&\!\! \xi 
   \end{array} \!\!\right]}
  }
             \lambdachi 
\end{displaymath}
\begin{displaymath}
{\cal V}=\frac{|e+m\tau|^2+\xi^2}{16 \pi \tau_2}
\end{displaymath}
where $\tau=\tau_1+i\tau_2$ is now a function of $a$, and we have
$\langle \tau \rangle = \frac{\theta}{2\pi}+i\frac{4\pi}{g^2}$ in terms
of the $\theta$-angle and coupling constant $g$.

\section{Partial Breaking of Supersymmetry}

We now turn to the minimization of ${\cal V}$. It has a stationary point at 
$\langle \tau'(a) \rangle =0$, but that corresponds to a saddle point 
and therefore does not give rise to a vacuum. The central charge 
matrix of eq. (\ref{al}) was computed in \cite{FGP2}:
\begin{equation}
C^{AB}=4\vec{\sigma}^{AB}\cdot(\vec{e}\times\vec{m})=-4 \xi m 
{ {\footnotesize \left[\!\!\begin{array}{cc} 0 & 1 \\ 1 & 0 
    \end{array}\!\!\right]}
  }\, .
\end{equation}
We distinguish between the following cases:

$\bullet \; \xi m \neq 0$ : the supersymmetric current algebra is modified
and therefore a partial breaking can occur in this case.  
Minimizing ${\cal V}$ with respect to $\langle \tau_1 \rangle$ 
and $\langle \tau_2 \rangle$ gives a vacuum at
\begin{equation}
\langle \tau_1 \rangle =-\frac{e}{m} \; , \;  \langle \tau_2 
\rangle = \left| \frac{\xi}{m} \right|\, ,
\end{equation}
from which a vacuum expectation $\langle a \rangle$ can be found. At
this point of 
the moduli space, one finds the following spectrum
\begin{equation}
\left(a, \frac{\lambda-s\chi}{\sqrt{2}}\right) \mbox{ of mass  } 
\frac{s m^2 }{2 \xi}\langle \tau' \rangle\, ,
\end{equation}
\begin{equation}
\left( A_\mu , \frac{\lambda+s\chi}{\sqrt{2}}\right) \mbox{ massless, with
  $s=\mbox{sign}(\xi m)$}\, ,
\end{equation}
which suggests that we end up with an $N=1$ vacuum. This can be checked
on the SUSY transformation of the Goldstino $(\lambda+s\chi)/\sqrt2$:
\begin{displaymath} 
\delta \left( \frac{\lambda +s\chi}{\sqrt{2}} \right) = 
im(s\epsilon^1-\epsilon^2 )\, ,\ \delta \left(\frac{\lambda -s\chi}{\sqrt{2}} \right) = 0\, ,
\end{displaymath}
where $\epsilon^{1,2}$ are the two infinitesimal supersymmetry 
parameters. Therefore, partial $N=2 \rightarrow N=1$ 
supersymmetry breaking does occur and we get here a massless $U(1)$ 
vector multiplet plus a massive chiral multiplet. Notice that, 
at any other point of the moduli space, the two 
supersymmetries are non-linearly realized and that only at the minimum of 
${\cal V}$, we get a combination $sQ^1-Q^2$ of the initial generators which is 
restored. Also, $\langle {\cal V} \rangle = |\xi m |/8\pi >0$ at the minimum but one 
could have chosen the parameters $\vec{p}$ in order to shift the scalar
potential to a zero vacuum energy (see eq. (\ref{ltot})).

$\bullet \; m\neq 0, \xi  =0$ :  in this case, we deal with the usual 
supersymmetric algebra. The previous minimum is now at
\begin{equation}
\langle \tau_1 \rangle =-\frac{e}{m} \; ,  \; \langle \tau_2 
\rangle = 0 \; .
\end{equation}
Since $4\pi{\cal L}_{\mbox{\em tot}}=-\langle \tau_2 \rangle \partial^\mu a 
\partial_\mu \bar{a} +\cdots$, we are sent to a point $\langle a 
\rangle$ where the K{\"a}hler metric $K_{a\bar{a}} \equiv 
\langle \tau_2 \rangle$ is singular. Several interpretations are 
then possible, depending on the underlying theory.

Firstly, if the theory is not asymptotically free, $g(\langle a 
\rangle)\rightarrow \infty$ and $\langle \tau_2 \rangle 
\rightarrow 0$ when $\langle a \rangle \rightarrow \infty$. 
In that case, the theory has a run away behavior. However, such an 
eventuality should not occur in a consistent underlying theory. 

Secondly, the singularity could be due to the appearance of new 
massless states at $\langle a\rangle$. For a non-Abelian 
underlying theory, these states cannot be vector multiplets because 
an enhanced gauge symmetry would be incompatible with the 
presence of Fayet--Iliopoulos terms.  However, in that case, dyonic 
states can be present in the theory and therefore we suppose that 
one of them becomes massless at this point. Let $(m_0,e_0)$ denote its 
magnetic and electric charges. In order to include it explicitly in our 
effective theory, we have to perform an electric--magnetic duality 
transformation which replaces the original photon $A_\mu$ by a 
``dyonic'' one which couples locally to the dyon. Let us define
\begin{equation}
{\footnotesize
    \left[\!\!\begin{array}{c} \tilde{F}'(\tilde{a})\\ \tilde{a}
    \end{array} \!\!\right]}~:={\footnotesize
    \left[\!\!\begin{array}{cc}\alpha&\!\!\beta\\ \gamma&\!\! \delta
    \end{array} \!\!\right]}
{\footnotesize \left[\!\!\begin{array}{c} F'(a)\\ a \end{array}
    \!\!\right]}\, , 
\end{equation}
where the matrix is of determinant one and chosen so that the BPS 
mass formula for the dyon takes the form \cite{SW} 
\begin{equation}
m = \sqrt{2} \left| (m_0 \delta-e_0 \gamma~=0, \tilde{q}){\footnotesize
    \left[\!\!\begin{array}{c} \tilde{F}'(\tilde{a})\\ \tilde{a}
    \end{array} \!\!\right]} \right| \, ,
\label{mass}
\end{equation}
where $\tilde{q}=-m_0 \beta+e_0 \alpha$.
Now, since 
\begin{equation}
\tilde{\tau}=\frac{\alpha \tau + \beta}{\gamma \tau + \delta} 
\sim -\frac{i}{\pi} \ln \tilde{a} \rightarrow \infty
\end{equation}
as $\tilde{a}\rightarrow 0$ i.e. $\tau \rightarrow -\frac{e}{m}$, from 
the vanishing of the denominator of $\tilde{\tau}$, one deduces that 
$(m,e)$ is orthogonal to $(\delta, -\gamma)$. Moreover, from eq. 
(\ref{mass}), we have $(\delta, -\gamma)$ orthogonal to $(m_0,e_0)$ 
and therefore it exists a constant $c$ such that
\begin{equation}
(m,e)=c(m_0,e_0) \, .
\label{c}
\end{equation}
The charges of the dyon are now $(0,\tilde{q})$ 
with respect to $\tilde{A}_\mu$. Performing the symplectic 
transformation on the Lagrangian ${\cal L}_{tot}$ in
eq. (\ref{Ltot1}) and using eq. (\ref{c}), one finds 
\begin{equation}
S_{\mbox{\em tot}} = \frac{\mbox{Im}}{4\pi}\int d^4x  [\tilde{{\cal L}}^1_0
+\tilde{{\cal L}}^1_{F.I.}]\, , \, \mbox{where}
\end{equation}
\begin{displaymath}
\tilde{{\cal L}}^1_0  =  \int d^2 \theta d^2\bar{\theta} 
\tilde{F}'(\tilde{A}) \bar{\tilde{A}}+\int d^2 \theta \frac{1}{2} 
\tilde{\tau} \tilde{W}^\alpha \tilde{W}_\alpha \, , 
\end{displaymath}
\begin{displaymath}
\tilde{{\cal L}}^1_{F.I.}= i \int d^2\theta \;
c\tilde{q}\tilde{A} \, ,
\end{displaymath}
to which one must add the coupling of $\tilde{A}_\mu$ to the dyon 
$(\tilde{\Phi}_+, \tilde{\Phi}_-)$  where
$\tilde{\Phi}_\pm$ are two $N=1$ chiral 
superfields. For simplicity, we choose the hyperk{\"a}hler manifold of 
the hypermultiplet to be flat:
\begin{eqnarray}
S_{\mbox{\em dyon}} \!\!\!\!&=&\!\!\!\! {1\over 4\pi}\int d^4x d^2 \theta d^2 \bar{\theta} 
\sum_{\sigma=\pm} \bar{\tilde{\Phi}}_\sigma e^{\sigma\tilde{q}\tilde{V}}
\tilde{\Phi}_\sigma \nonumber \\ 
& & +{\mbox{Im}\over4\pi}\left[ \sqrt2 \tilde{q} \int d^4x d^2\theta \tilde{A} \tilde{\Phi}_+
\tilde{\Phi}_-\right] \, .
\label{Shyp} 
\end{eqnarray}
Minimizing the new scalar potential, one finds that at 
the singular vacuum $\langle \tilde{a} \rangle =0$, the
$\tilde{\Phi}_\pm$ scalar VEV's $\langle \tilde{\phi}_\pm\rangle$ are such that
\begin{equation}
|\langle\tilde{\phi}_+\rangle|=|\langle\tilde{\phi}_-\rangle|\, , \
\langle\tilde{\phi}_+\rangle \langle\tilde{\phi}_-\rangle ={ic\over\sqrt2}\, . 
\end{equation}
We have a condensation of dyons which breaks the $U(1)$ gauge symmetry
in a zero energy vacuum. Since we deal with the usual supersymmetry 
algebra, this is enough to show that $N=2$ is unbroken. 
This phenomenon is similar to the one found in \cite{SW} where 
such a condensation happened in the case of an explicit breaking of 
$N=2$ to $N=1$. The spontaneous mechanism we have here is also similar to the 
effect induced by a generic superpotential near the conifold 
singularity of type II superstrings compactified on a Calabi-Yau 
threefold \cite{S}. In that case, the massless hypermultiplets are 
black holes which condense at the conifold points, which leads also to
new $N=2$ vacua. One way to generate such a superpotential is to give
a VEV to the 10-form \cite{SP}, which should correspond in 4
dimensions to a magnetic Fayet-Iliopoulos term.

$\bullet \; m=0, \xi \neq 0$ : this case is similar to the previous one 
except that the metric is now $\langle \tau_2 \rangle =+\infty$ 
at the minimum. A similar analysis shows that this can happen either
when the theory is asymptotically free, which gives rise to a run away
behavior, or if electric hypermultiplets become massless at this point.  

$\bullet \; m=\xi=0,\, e\neq 0$ : in that case, ${\cal V}=e^2/16\pi \tau_2 
\rightarrow 0$ when $\tau_2 \rightarrow +\infty$ and the conclusion of the
previous case are still valid here.

$\bullet \; m=\xi=e=0$ : $N=2$ and the $U(1)$ gauge symmetry are
not broken. 

\section{Gauge theory with matter}

Having discussed partial spontaneous SUSY breaking in a pure gauge theory,
we shall now consider extending it to a theory with charged matter, the 
simplest case being one hypermultiplet coupled to a $U(1)$ vector multiplet.
Surprisingly, we shall find that the theory does not admit any magnetic 
Fayet--Iliopoulos term anymore (while it still keeps the electric ones). 
There are already a few hints that could
prevent us from boldly extrapolating from the pure gauge theory. First,
Fayet--Iliopoulos terms are intimately related with expectation values of auxiliary fields,
whereas the hypermultiplet does not admit a (finite) off-shell formulation.
Second, existence of charged matter breaks electric--magnetic duality,
so we cannot generate a magnetic Fayet--Iliopoulos term from an electric one by a symplectic
transformation. Of course, we will not be able to rigorously prove
the inexistence of such a deformation, but we shall at least show why a
straightforward generalization fails.

Let us therefore consider the simplest case of hypermultiplet
$\Phi_\pm\equiv (\phi_\pm,\psi_\pm)$, described by a flat 
hyperk{\"a}hler manifold ${\bf R}^4$, and couple it with charge $q$ to a 
vector multiplet $a,A_\mu,\lambda,\chi$ with arbitrary prepotential
and Fayet-Iliopoulos terms. The Lagrangian takes the form~:
\begin{equation}
{\cal L}_{v+h}={\cal L}_{tot}+{\cal L}_{hyp}+{\mbox{Im}\over 2\pi}\int
d^2 \theta iM\Phi_+\Phi_- \, ,
\end{equation}
where $M$ is a complex mass term, ${\cal L}_{tot}$ is defined in
eq. (\ref{Ltot1}), and ${\cal L}_{hyp}$ is formally given by the
Lagrangian of the action in eq. (\ref{Shyp}) with the tildes removed.  

By construction,
we obtain a $N=1$, presumably $N=2$, supersymmetric Lagrangian, from which 
we can work out the scalar potential
and the fermion mass matrix by eliminating the auxiliary fields.
Aiming at proving its $N=2$ invariance, we display the result in a
$SU(2)_R$ invariant way, where $SU(2)_R$ is the automorphism group
of $N=2$, $D=4$ supersymmetry algebra~:
\begin{eqnarray}
{\cal V}\!\!\!\! &=&\!\!\!\! {\cal V}_{C} + {\cal V}_{NC}\, \mbox{, with}\label{pot}\\
{\cal V}_{C}\!\!\!\!&=&\!\!\!\!{1\over 4\pi}\left| {iq\over \sqrt2}a-M \right|^2
\left( |\phi_+|^2 + |\phi_-|^2 \right) \nonumber \\
         \!\!\!\!& &\!\!\!\!  + {1\over 16\pi\tau_2} | {q\vec{D}\over
             \sqrt2}-\vec{e}-\vec{m}\bar{\tau} | ^2 \nonumber \\
{\cal V}_{NC} \!\!\!\!&=&\!\!\!\! - {mq\sqrt2\over 8\pi} \; \mbox{Re}(
\phi_+ \phi_-)\, ,
\nonumber 
\end{eqnarray} 
\begin{eqnarray}
{\cal M}\!\!\!\! &=&\!\!\!\! {-i q \over 4\sqrt{2}\pi} 
            \lambdachitransp \vierbein \sigmathree \twopsi \nonumber \\
        \!\!\!\! & &\!\!\!\!  - {1\over8\pi}\!\left({iq\over\sqrt2} a - M \right)\!
         \twopsitransp \!\sigmathree\! \twopsi \nonumber \\
        \!\!\!\! & &\!\!\!\!  + {\tau'\over 32\pi\tau_2} 
             \lambdachitransp \left( {q\vec{D}\over\sqrt{2}}
                                     -\vec{e}-\vec{m}\bar{\tau}
                                   \right)\cdot \vec{\sigma} 
             \lambdachi \, , \nonumber
\end{eqnarray}             
where $SU(2)_R$ acts on the left of the hypermultiplet vierbein
${\cal U}=\vierbein$ and 
the gauginos $\lambdachi$, whereas the gauge group $U(1)$,
embedded in the direction $\sigma_3$ of the symplectic group $Sp(1)=SU(2)$, 
acts on the right of the vierbein and on the left of the hyperino doublet
$\twopsi$. 
$\vec{D}:={1\over2} \mbox{tr}\, {\cal U} \sigma_3 {\cal U}^+ \vec{\sigma}$ 
is the triholomorphic moment map for the hypermultiplets
(see \cite{boris} for a pedagogical review of 
hypermultiplet geometry).

In fact, it is not quite invariant under $SU(2)_R$, because of the second term 
${\cal V}_{NC}$ 
in the scalar potential, proportional to $D_1 m_2$~: this sheds some doubt
on the $N=2$ SUSY of the model. On the other hand, it may be shown that
the covariant piece ${\cal V}_{C}$ is precisely obtained when one covariantizes
the general potential of a rigid $N=2$ theory \cite{ABCdAFFM} with respect to
the symplectic duality group, i.e. by adjoining to the triholomorphic
moment map a dual 'magnetic' moment map to make a true symplectic section. 

In order to check $N=2$ SUSY, we have to work out 
the SUSY variations of the fields:
\begin{eqnarray*}
\delta a \!\!\!\!&=&\!\!\!\! -\sqrt{2}\; \lambdachitransp \twoeps  
\end{eqnarray*}
\begin{eqnarray*}
\delta\lambdachi\!\!\!\!&=&\!\!\!\! i \sqrt{2}\; \partial_\mu a \sigma^\mu \twoepsconj
  + F_{\mu\nu} \sigma^{\mu\nu} \twoeps \\
 \!\!\!\!& &\!\!\!\!- {i \over\sqrt{2}\tau_2} \left( {q\vec{D}\over\sqrt{2}}
                                   -\vec{e}-\vec{m}\bar{\tau} \right)\cdot 
    \vec{\sigma} \twoeps   \\
\!\!\!\!\! & &\!\!\!\!\! + { \sqrt{2} i \over 4 \tau_2}\!
    \left( \tau'\! \lambdachi\! \lambdachitransp
          \!+\! \bar{\tau}'\! \lambdachiconj \!\lambdachidag 
    \right) \!\!\twoeps  
\end{eqnarray*}
\begin{eqnarray*}
\delta\vierbein\!\!\!\! &=&\!\!\!\! -\sqrt{2} \twoeps \twopsitransp \nonumber\\
               \!\!\!\! & &\!\!\!\!  +\sqrt{2} \twoepsconj \twopsidag 
\end{eqnarray*}
\begin{eqnarray*}
\delta\twopsitransp \!\!\!\!&=&\!\!\!\! -i \sqrt{2} \sigma^\mu
                 \twoepsdag \nabla_\mu \vierbein \nonumber\\
              \!\!\!\! & &\!\!\!\!+\sqrt{2}\left({i q\over\sqrt{2}}
                 \bar{a} + \bar{M}\right) \\
               \!\!\!\!& &\ \ \ \twoepstransp\vierbein\sigmathree \,
                 .
\end{eqnarray*}
These variations are imposed to us by the requirement that they should
reduce to the usual variations when restricting to $N=1$ SUSY ($\epsilon^2=0$),
to the $N=2$ variations when one forgets the hypermultiplets, and by $SU(2)_R$ 
covariance. Alternatively, they can be obtained by symplectic covariantization 
of the general SUSY variations of \cite{ABCdAFFM}. Provisionally forgetting 
the non-covariant term, we can now 
evaluate up to three fermion terms the holomorphic
($\bar{\epsilon}^1=\bar{\epsilon}^2=0$) SUSY variation of the 
Lagrangian~:
\begin{eqnarray*}
\phantom{.}& &\!\!\!\!\!\!\!\!\!\!\!\!\!\delta\left({\cal M + V}_{C}\right) = \\
& &{i q \over 2} \twopsitransp \sigmathree
     \vierbeinbar \vec{m}.\vec{\sigma} \twoeps \\
& &= - \delta {\cal V}_{IM}\, \mbox{, with}\\
& &\!\!\!\!\!\!\!\!\!\!\!\!\!{\cal V}_{IM} = -{i q\over 8\pi\sqrt{2}} \vec{D}\cdot\vec{m} = 
  -i{\sqrt{2}\over8\pi} mq\; \mbox{Im}(\phi_+ \phi_-)\, .
\end{eqnarray*} 
So we see that in order to compensate this variation we would have to add
to the covariant Lagrangian ${\cal V}_{C}$ a piece 
${\cal V}_{IM}$, covariant but purely imaginar. Compensating the
antiholomorphic variation instead would require adding the complex 
conjugate piece $-{\cal V}_{IM}$, so there is really no way to 
implement $N=2$ SUSY ! Taking into account ${\cal V}_{NC}$, $N=1$
supersymmetry is explicitly checked, since
${\cal V}_{IM}$ and ${\cal V}_{NC}$ have the same holomorphic part,
while the $N=1$ SUSY transformation $\epsilon^2=0$ involves only the
holomorphic derivatives of the potential. Lastly, if one forgets both
the non-covariant and imaginary pieces, we find that the naive covariantized
version of the general N=2 Lagrangian is not even supersymmetric !

In case the arguments presented above should not be compelling enough,
it is still possible to evaluate the spectrum in vacua of the theory
with hypermultiplets and determine from the residual SUSY and the number
of Goldstinos the original SUSY, and the result confirms that the
Lagrangian obtained by minimally coupling the hypers to the pure 
Fayet-Iliopoulos deformed gauge theory does have N=1 SUSY only.

\vspace{.4cm}
\noindent{\bf Acknowledgements~:} 
we are grateful to I. Antoniadis for very useful conversations.

\end{document}